\documentclass[]{article}
\usepackage{graphicx}                    
\usepackage[square]{natbib}
\begin{document}
\title{Spatial \& Temporal Characteristics of Ha flares during the period 1975--2002 (comparison with SXR flares)}

\author{E. Gini, J. Polygiannakis\footnote{Deceased},\\ A. Hillaris, P. Preka--Papadema,X. Moussas\\Section of Astrophysics, Astronomy and Mechanics, 
Department of Physics,\\ {\emph{University of Athens, 15784 Panepistimiopolis Zografos, Athens, Greece}}}
\maketitle
\begin{abstract}
Although the energetic phenomena of the Sun (flares, coronal mass injections etc.) exhibit intermittent stochastic behavior in their rate of occurrence, they are well correlated to the variations of the solar cycle. In this work we study the spatial and temporal characteristics of transient solar activity in an attempt to statistically interpret the evolution of these phenomena through the solar cycle, in terms of the self-organized criticality theory.  
\end{abstract}

\section{Introduction}
\begin{figure}
  \includegraphics[width=\textwidth]{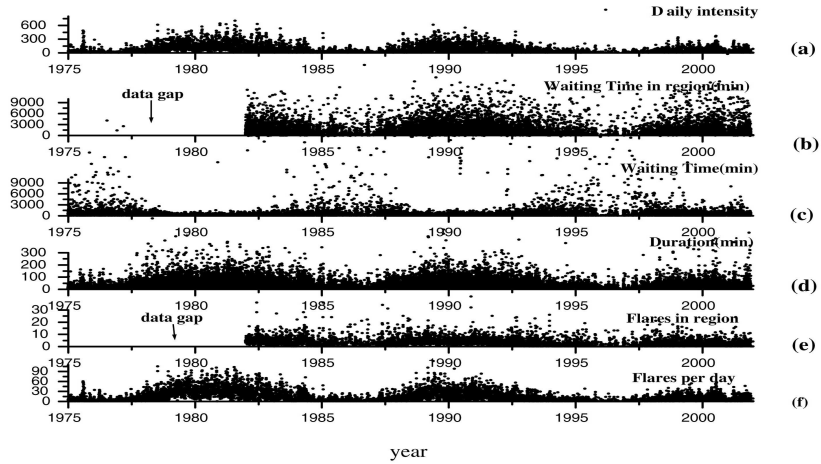}
\caption{Variation of Flare Parameters with the Solar Cycle: (a) Daily Intensity (SXR);
it has been included for comparison as it marks the Solar Cycle in exactly the same way as
the Sunspot Number. (b) Waiting Time between Successive Ha flares in the same active region. 
The data gap is due to the fact that AR number fro Ha flare has not been included in SGR since 1982.
(c)Waiting Time between Successive Ha flares (d)Ha Flare Duration (e) Number of flares
per day in the same active region. There is a data gap from 1975 to 1982
(f). Number of Flares Daily.}
\label{Figure01}
\end{figure}
\begin{figure}
\begin{minipage}[t]{6cm}
\begin{center}
\includegraphics[width=6cm]{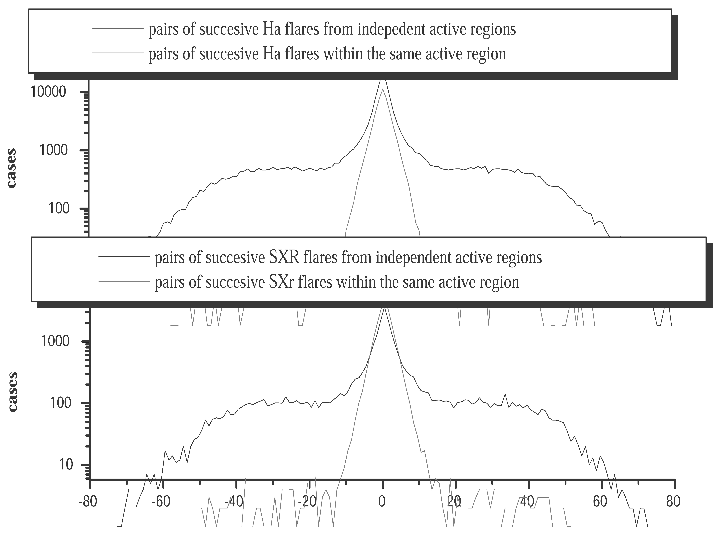}
\end{center}
\end{minipage}
\hfill
\begin{minipage}[t]{6cm}
\begin{center}
\includegraphics[width=6cm]{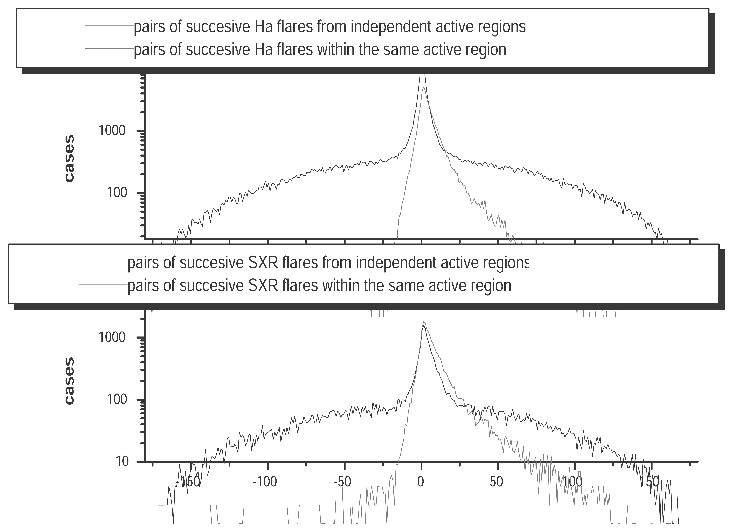}
\end{center}
\end{minipage}
\caption{Angular Separation of Successive Flares; Upper panels refer to Ha flares
lower to SXR: Left: Separation in Lattitude
Right: Separation in Longitude. The peak corresponds to events within the same
active region, the plateau to flares on the disk in general.}
\label{Figure04}
\end{figure}
\begin{figure}
\begin{minipage}[t]{6cm}
\begin{center}
\includegraphics[width=6cm]{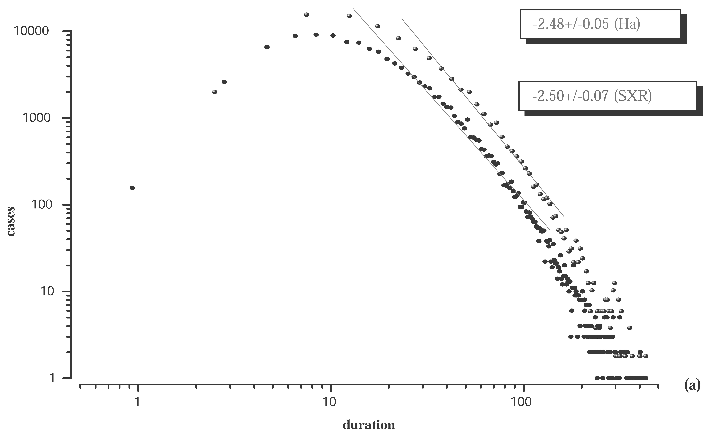}
\end{center}
\end{minipage}
\hfill
\begin{minipage}[t]{6cm}
\begin{center}
\includegraphics[width=6cm]{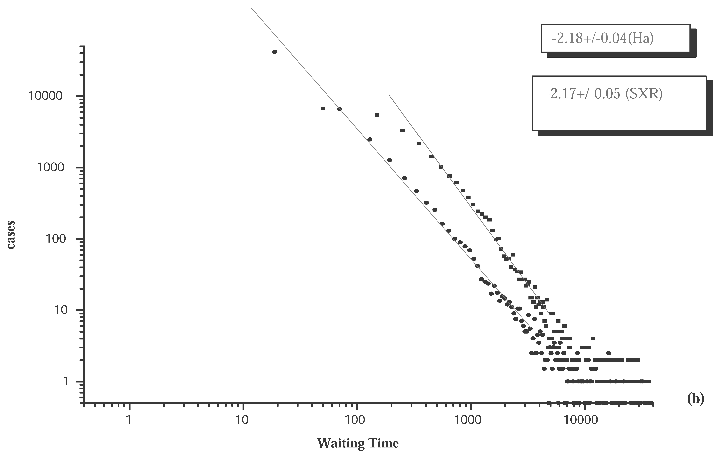}
\end{center}
\end{minipage}
\begin{minipage}[t]{6cm}
\begin{center}
\includegraphics[width=6cm]{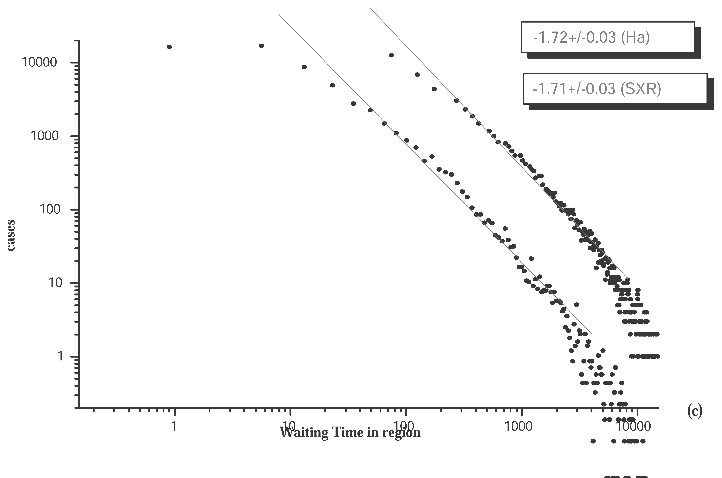}
\end{center}
\end{minipage}
\hfill
\begin{minipage}[t]{6cm}
\begin{center}
\includegraphics[width=6cm]{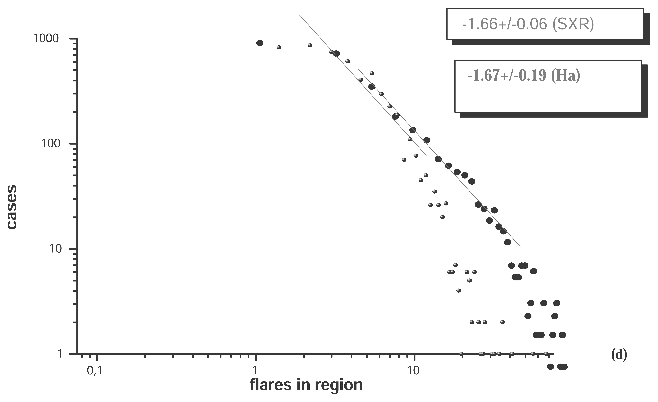}
\end{center}
\end{minipage}
\caption{Ha Flare Statistics Compared to SXR: Top Left: Flare Duration ($D$);
Bottom Left: Interval Between Successive Flares in the same Active Region ($\Delta T_{AR}$)
Top Right:Interval Between Successive Flares $\Delta T$; 
Bottom Right: Number of Flares per Active Region ($NF$).}
\label{Figure05}
\end{figure}
A statistics description of solar energetic phenomena, furnishes much information 
despite the inherently complex and, hence, incompetely undestood processes of 
magnetic energy storage and release which drives them. 

The \emph{Self Organised Criticality (SOC)} constitutes such a promising approach 
(cf. \cite{Bak} for a review). In the SOC paradigm of sand pile model periods of stasis 
are interrupted by intermittent sand slides, or avalanches (a domino effect) and the number 
of avalanches $N(s)$ of size, s, is given by power law distribution $N(s) \approx s^a$, 
where, a, is the power-law exponent, characteristic of the particular physical process. 
Power--law size distributions can be reproduced computationally as the output of cellular automata models, 
in which energy is added, cascaded and released as a collective phenomenon on a grid. 

Solar flares exemplify \emph{Self Organised Criticality (SOC)} because 
they represent an almost explosive release of previously stored magnetic
energy (\cite{Parker88}, \cite{Parker89}); this has been duly demonstrated in 
\cite{Lu91}, \cite{Lu93},  \cite{Georgoulis}, \cite{Anastasiadis}, \cite{Mac},
\cite{Vlahos} etc. They have simulated SOC by means of \emph{cellular automata (CA)} models, 
which reproduced the power law behavior in the distribution functions of the total
energy, the peak luminosity and the duration of the flares. 
The results were observationally confirmed as regards the distribution of the HXR 
peak flux, total energy associated, flare duration (cf. \cite{Veronig}, \cite{Crosby},
\cite{Crosby93}, \cite{Crosby99}) etc.

The flaring process, has an intermittent and, apparently, random character modelled well by the 
SOC. The rate of magnetic energy emergence, on the other hand, follows in general, the solar
cyle evolution and exhibits recuring patterns as regards the appearance on the disk (butterfly);
it provides, thus, a quasi--periodic component in the flaring activity as is represents the rate
of accumulation of the energy which, in turn, is explosively released in flares.

In this report we expand on a previous work (\cite{Polygiannakis}) on the spatial and temporal distribution of 
energetic phenomena on the sun. This work was based on GOES SXR data in the 1975--2002 period; in this
report we extend our analysis, including the SGD Ha flares in the same period. 

\section{DATA ANALYSIS AND RESULTS}
From the \emph{Solar Geophysical Data  (SGD)} we have analysed and compared SXR and Ha flare
parameters in the 1975--2002 period. We only used events with recorded time of start. 
Since the data cover more than one solar cycle (in particular cycles 21, 22 and partly
23) we were able to study their dependence on the to the cycle's phase .

In figure \ref{Figure01} we note that the total intensity\footnote{Which follows the Sunspot Number}, 
(\ref{Figure01}(a)), is positively correlated with the daily number of events,
(\ref{Figure01}(f)), the number of events in an active region (\ref{Figure01}(e)), 
the event duration, (\ref{Figure01}(d)), and the waiting time between
events within the same region, (\ref{Figure01}(b)), 
while is anticorrelated with the waiting time between events
regardless of region, (\ref{Figure01}(c)). 

The distributions of event duration ($D$) (figure \ref{Figure05}(a)), 
waiting times between successive events on disk ($\Delta T$), 
(figure \ref{Figure05}(b)) and in the same active region ($\Delta T_{AR}$), 
(figure \ref{Figure05}(c)), and number of flares in the same region ($NF_{AR}$), 
(figure \ref{Figure05}(d)) exhibit power--law tails. This reflects the wide range of energy release
time scales in flares. Althougth the non--Poisson, power--law distributions 
of the intervals between successive events have been challenged at times
(cf. \cite{Crosby99} and references within) they appear to underline  that the individual 
flares are correlated. Long range time correlations are
expected since the active regions are magnetically connected and the flares represent a 
redistribution of the magnetic energy and alteration of the magnetic topology by reconnection.
This affects the lifetime of neighbouring structures and often triggers new flares, thus the trend for
flares to appear clustered in time.
 
The spatial distribution of successive events on the solar disk was also studied; 
we derived the distribution of angular separation of pairs of successive events.
Most occur in the same active region, with exponentially decaying probability distribution. 
The distribution of separation in longitude is asymmetric
because of the solar differential rotation (cf. figure \ref{Figure04}). 

\section{CONCLUSIONS}
The conclusions support and reaffirm previous results of \cite{Polygiannakis}. In particular:
\begin{enumerate}
\item{The daily number of SXR and Ha flares in same active region, event duration and waiting time in 
same active region follow the general solar cycle. The same solar cycle dependence was derived for the flares
on the whole disk regardless of active region. What constitutes a marked exception to this trend is the
waiting time between successive flares on disk; this was found anticorrelated to the solar cycle.}

\item{ The distributions N, of event duration (D),waiting time between successive 
events\footnote{on disk or within each active region; the latter has been subscripted with AR} 
($\Delta T$, $\Delta T_{AR}$ and number of flares occurring in the same active region $NF_{AR}$ 
exhibit power--law parts over a broad range of parameters. The exponents calculated are: 
\begin{itemize}
\item{$N \propto D^{-2.48 \pm 0.05}$ for Ha compared to  $N \propto D^{-2.50 \pm 0.07}$ 
for SXR\footnote{In \cite{Veronig} the SXR duration power--law differs a little, 
being $N \propto D^{-2.93 \pm 0.12}$}}
\item{$N \propto \Delta T^{-2.18 \pm 0.04}$ for the waiting time between Ha flares compared 
to $N \propto \Delta T^{-2.17\pm 0.05}$ for SXR}
\item{$N \propto \Delta {T_{AR}}^{-1.72 \pm 0.03}$ for the waiting time between Ha flares in the same AR 
compared to  $N \propto \Delta {T_{AR}}^{-1.71 \pm 0.03}$ for SXR}
\item{$N \propto {NF_{AR}}^{-1.67 \pm 0.19}$ for the number of Ha flares within the same 
active region compared to $N \propto {NF_{AR}}^{-1.66 \pm 0.06}$ for SXR.} 
\end{itemize}
The power--law exponents are, within the margin of statistical error, the same for Ha and SXR. They confirm thus
the previous conclusions by \cite{Polygiannakis} in support of the hypothesis of 
meta-stable, SOC state of magnetic structures of the Sun. 
The power--law distributions of waiting times indicate that each event triggers others 
due to the spatial interconnection of the erupting magnetic structures.}

\item{The study of the angular separation between successive events shows that mostly, energetic events
tend to cluster within the same active region.}

\end{enumerate}

{\emph{This work is dedicated to the departed member of the ARTEMIS--IV Group, Dr. John Polygiannakis}}


\begin{thebibliography}{}

\bibitem{Anastasiadis}
A. Anastasiadis, \emph{First School on Physics and Technology of Fusion Proceedings, (Eds. A. Grekos)} 115--129, (2003).

\bibitem{Bak}
P., Bak, \emph{How nature works, the science of self--organized criticality, Copernicus, 
Springer--Verlag, New York}, (1996).

\bibitem{Crosby}
N.~B., Crosby, M.~J., Aschwanden, B.~R., Dennis, \emph{Sol.Phys.}, \textbf{143}, 275--299, (1993).

\bibitem{Crosby93}
N.~B., Crosby, M.~J., Aschwanden, B.~R., Dennis, \emph{Advances in Space Research}, \textbf{13}, 179--182, (1993).

\bibitem{Crosby99}
N.~B., Crosby, M.~K., Georgoulis, , N., Vilmer, \emph{ESA SP} \textbf{446}, 247--250, (1999).

\bibitem{Georgoulis}
M.~K., Georgoulis, L., Vlahos, \emph{Astron. Astrophys.}, \textbf{336}, 721--734, (1998).

\bibitem{Lu91}
E.~T., Lu and R.~J, Hamilton, \emph{Astrophysical Journal}, \textbf{380}, L89--L92 (1991).

\bibitem{Lu93}
E.~T., Lu and R.~J, Hamilton, J.~M., Mc Tiernan and K.~R., 
Bromund, \emph{Astrophysical Journal}, \textbf{412} 841-852 (1993).

\bibitem{Mac}
A.~L., MacKinnon, K.~P., Macpherson, L., Vlahos, \emph{Astron. Astrophys.}, \textbf{310}, L9--L12, (1996).

\bibitem{Parker88}
E.~N., Parker, \emph{Astrophys. J}, \textbf{330}, 474--479, (1988).

\bibitem{Parker89}
E.~N., Parker, \emph{Sol. Phys.}, \textbf{121}, 271-288, (1989).

\bibitem{Polygiannakis}
J.~M., Polygiannakis, A., Nikolopoulou, P., Preka-Papadema, X., Moussas,
A., Hillaris,  \emph{ESA SP} \textbf{505}, 541--544, {2002}.

\bibitem{Veronig}
A., Veronig, M., Temmer, A.,  Hanslmeier, W.,  Otruba, and M.,  Messerotti, \emph{Astron. Astrophys.}, \textbf{382}, 
1070--1080 (2002)

\bibitem{Vlahos}
L., Vlahos, M.~K., Georgoulis, 
R., Kluiving and P., Paschos, \emph{Astron. Astrophys.}, \textbf{299}, 897--911 (1995).

\end{thebibliography}
\end{document}